\documentclass{article}

    \usepackage[preprint]{neurips_2025}

\usepackage[utf8]{inputenc} % allow utf-8 input
\usepackage[T1]{fontenc}    % use 8-bit T1 fonts
\usepackage{hyperref}       % hyperlinks
\usepackage{url}            % simple URL typesetting
\usepackage{booktabs}       % professional-quality tables
\usepackage{amsfonts}       % blackboard math symbols
\usepackage{nicefrac}       % compact symbols for 1/2, etc.
\usepackage{microtype}      % microtypography
\usepackage{xcolor}         % colors
\usepackage{graphicx}
\usepackage{caption}
\usepackage{enumitem}
\usepackage{wrapfig}
\setlist{itemsep=0em, topsep=0em, parsep=0.5em, partopsep=0em, leftmargin=2em}
\usepackage{makecell}

\newcommand{\reaper}{\texttt{REAPER}}

\title{WildFX: A DAW-Powered Pipeline for In-the-Wild Audio FX Graph Modeling}

\author{%
  Qihui Yang\\
  UC San Diego \\
  \texttt{qiy009@ucsd.edu} \\
  \And
  Taylor Berg-Kirkpatrick\\
  UC San Diego \\
  \texttt{tberg@ucsd.edu} \\
  \And
  Julian McAuley\\
  UC San Diego \\
  \texttt{jmcauley@ucsd.edu} \\
  \And
  Zachary Novack\\
  UC San Diego \\
  \texttt{znovack@ucsd.edu}\\
}

\begin{document}

\maketitle

\begin{abstract}
Despite rapid progress in end-to-end AI music generation, 
AI-driven modeling of professional Digital Signal Processing (DSP) workflows remains challenging. 
In particular, while there is growing interest in neural black-box modeling of audio effect graphs (e.g. reverb, compression, equalization), 
AI-based approaches struggle to replicate the nuanced signal flow and parameter interactions used in professional workflows. Existing differentiable plugin approaches often diverge from real-world tools, exhibiting inferior performance relative to simplified neural controllers under equivalent computational constraints. We introduce \textbf{WildFX}, a pipeline containerized with Docker for generating multi-track audio mixing datasets with rich effect graphs, powered by a professional Digital Audio Workstation (DAW) backend. \textbf{WildFX} supports seamless integration of cross-platform commercial plugins or any plugins in the wild, in VST/VST3/LV2/CLAP formats, enabling structural complexity (e.g., sidechains, crossovers) and achieving efficient parallelized processing. A minimalist metadata interface simplifies project/plugin configuration. Experiments demonstrate the pipeline’s validity through blind estimation of mixing graphs, plugin/gain parameters, and its ability to bridge AI research with practical DSP demands. The code is available on: \href{https://github.com/IsaacYQH/WildFX}{https://github.com/IsaacYQH/WildFX}
\end{abstract}

\section{Introduction}

Recent advances in large-scale music generation \cite{agostinelli2023musiclm, donahue2023singsong, Novack2025Presto, forsgren2022riffusion, yuan2025yue} have delivered remarkable end-to-end systems, from individual instrument sample generators \cite{Novack2025Fast, nercessian2023instrumentgen} to complete text-to-song models \cite{yuan2025yue}. From research-based open models like Stable Audio Open \cite{evans2024open} and YuE~\cite{yuan2025yue} to commercial systems like Suno and Udio, such generative models
have demonstrated the potential for AI to revolutionize creative audio workflows.
Despite these achievements in generative modeling, there remains a significant disconnect between modern AIxMusic research and the professional Digital Signal Processing (DSP) tools that form the backbone of modern music production, where full-stack song generation methods lack the flexibility and integration into professional musical workflows.

Such disconnect has motivated a large endeavor of research into neural audio effect modeling (NeuralAFx), which can act as a bridge between AI capabilities and traditional DSP techniques \cite{martinez2020deep, hawley2019signaltrain, steinmetz2021automatic, steinmetz2022style, steinmetz2021efficient}. Such research has spanned from simple discriminative tasks like plugin identification \cite{comunita2020guitar}, to parameter estimation \cite{mitcheltree2023modulation}, full AFx graph estimation \cite{lee_blind_2023}, effect modeling \cite{comunita_differentiable_2025, yu2023singing}, style transfer \cite{steinmetz_st-ito_2024}, and broader AI-assisted mixing and mastering \cite{koo2023music}. In particular, much of this research is guided by the principle of building \emph{differentiable} equivalents of traditional DSP modules, thus enabling partial- or fully-neural AFx plugins and analysis tools. Such approaches have grown in popularity in recent years, with even some limited deployment in commercial music plugins like \textit{Neural DSP}.
% Researchers have explored neural approaches to modeling individual effects like reverberation (Steinmetz et al., 2021), dynamic range compression (Steinmetz \& Reiss, 2021), and equalization (Colonel et al., 2022), as well as more complex multi-effect processing chains (Steinmetz et al., 2022; Hawley et al., 2022).
% However, a consistent emphasis on differentiability in these approaches has led to simplified approximations of professional tools, often evaluated on constrained tasks that poorly reflect real-world production scenarios.

However, this focus on differentiable architectures, and direct capacity to interface with Python more broadly, has created a widening gap between academic research and industry practice. Professional audio engineers and music producers do not work with Python-based differentiable modules; instead, they rely on complex commercial plugin ecosystems within Digital Audio Workstations (DAWs) such as Ableton or Logic Pro. 
% These plugins represent decades of industry expertise and often feature intricate parameter interactions, proprietary algorithms, and complex hand-crafted signal routing capabilities. 
Additionally, there is little evidence that fully neural AFx modules are comparable in performance to professional-grade plugins, with even simple neural baselines (i.e.~learning \emph{parameters} of a plugin rather than directly modeling the plugin itself) performing similarly \cite{steinmetz2022style, steinmetz2023high}. Because of this focus, most existing packages for  Neural AFx research involve either  differentiable modules or simple linux-based effects~\cite{engel2020ddsp, comunita_nablafx_2025, steinmetz2022style, uzrad2024diffmoog, lee2024grafx, yeh2024pyneuralfx}; this thus further widens the gap between research and practice, as even tasks that feasibly \emph{could} be applied to general commercial plugins (such as parameter estimation, graph learning, or simple discriminative problems) are tested on the limited and underperforment AFx modules supported in python.

We argue that to advance Neural AFx in ways that meaningfully impact professional audio production, research must engage directly with the tools already used by industry professionals. To address this need, we introduce \textbf{WildFX}, \textit{the first comprehensive end-to-end pipeline} (to the best of our knowledge) for interfacing with and generating multitrack music datasets with heterogeneous AFx graphs derived from universal plugins including real, commercial plugin chains using Python. WildFX is containerized with Docker, enabling efficient execution of a professional DAW backend (specifically \reaper{}) on Linux-based research systems—environments where audio production software typically does not run natively. This architecture supports seamless integration of arbitrary commercial plugins across multiple formats (VST/VST3/LV2/CLAP), allowing researchers to capture the full complexity of professional audio processing, including advanced routing schemes such as sidechaining and multiband processing. While the underlying design supports various forms of control signal routing, we refer to these uniformly as “sidechain” connections throughout this work for clarity.

The resulting datasets support a wide range of machine learning tasks, including plugin classification, parameter estimation, grey-box modeling, mixing graph inference, and musically informed source separation. Moreover, WildFX enables principled data augmentation for the music domain, where high-quality datasets are often limited due to copyright and production constraints. Its end-to-end architecture allows for plugin-driven audio transformations that better reflect real-world workflows. Unlike conventional augmentation strategies used in speech or general audio, plugin-based transformations align more closely with the practices of professional music production.

% [The dataset generation schema could add fuel to many ML tasks including plugin classification and parameter estimation, grey-box AFx modeling, mixing graph structure estimation, complex source separation and so on. we also point out that it is also good for music data augmentation. For LLM and Vision models, the major improvements are made by the success of scaling up the strong enough models. But for audio and music, especially for music, the amount of high-qualify data is limited for scaling up foundation models. That's because the nature of music data, limitation of copyrights. And compared to vision, text, speech, music is the hardest to generate without profession knowledge $\to$ synthetic data and data augmentation. Directly applying data augmentation from audio or speech may not make sense for music data. $+$ music producers would use audio plugins in their workflow $\to$ using audio plugins to do data augmentation makes more sense to music data.] 

The WildFX pipeline offers several key advantages over existing approaches. First, it enables the creation of datasets that reflect actual industry practices rather than simplified approximations. Second, it provides a minimalist metadata interface that simplifies project and plugin configuration, reducing the technical barriers to working with complex audio processing chains. Finally, it achieves efficient parallelized processing, making it practical for generating large-scale datasets necessary for training robust neural models.
To demonstrate the pipeline's effectiveness, we conducted experiments on blind estimation of mixing graphs, including detection of plugin types and parameter settings.
% [RESULTS SUMMARY] 
Our experiments show competitive results despite challenging settings, demonstrating that WildFX can generate realistic data suitable for training and evaluating neural audio systems.
 Our results highlight WildFX's potential to bridge the gap between AI research and practical DSP demands, enabling more ecologically valid neural modeling of audio processing workflows used by professionals in the field.

\vspace{-0.5em}
\section{Related Works}

\subsection{Neural Audio Effect Modeling (NeuralAFx)}

 NeuralAFx has seen significant growth in recent years, spanning multiple tasks and applications. Early work in this domain focused on basic classification tasks such as 
identifying guitar amplifier and pedal types from processed audio \cite{comunita2020guitar}. More popularly includes the subfield of \emph{parameter} estimation, wherein one uses gradient-based methods to learn the parameters of a normal AFx module, such as Low-Frequency Oscillators~\cite{mitcheltree2023modulation} or compressors~\cite{hawley2019signaltrain}.
% Comunità et al. \cite{comunita2020guitar} demonstrated successful identification of guitar amplifier and pedal types from processed audio, establishing the foundation for more complex discriminative tasks in audio processing. 
% Building upon these fundamentals, parameter estimation emerged as a natural extension. Mitcheltree et al. \cite{mitcheltree2023modulation} developed techniques for accurately estimating modulation effect parameters from processed audio, while Hawley et al. \cite{hawley2019signaltrain} presented SignalTrain, a deep learning approach for profiling audio compressors. These approaches laid groundwork for more comprehensive modeling tasks.
More recently, researchers have tackled the challenge of full audio effect graph estimation~\cite{lee_blind_2023, steinmetz_st-ito_2024}, wherein one infers the parameters of each AFx model \emph{and} their ordering with the audio processing graph.
% proposed methods for blindly estimating complete processing chains, inferring both the types of effects and their sequential ordering from processed audio examples. This represents a significant step toward understanding complex audio production workflows through machine learning.
Beyond these tasks, substantial effort has been directed toward black-box neural modeling of audio effects~\cite{martinez2020deep, steinmetz2021efficient, yu2023singing, comunita_differentiable_2025}.
% Martínez Ramírez and Reiss \cite{martinez2020deep} provided a comprehensive overview of deep learning approaches for black-box modeling of audio effects. Steinmetz et al. \cite{steinmetz2021efficient} demonstrated efficient neural networks for real-time analog audio effect modeling, while Yu et al. \cite{yu2023singing} explored specialized effect modeling for vocal processing.
% The field has also expanded into creative applications, with Steinmetz et al. \cite{steinmetz_st-ito_2024} developing neural approaches for audio effect style transfer, allowing the characteristics of one processing chain to be applied to new audio content. Comunità et al. \cite{comunita_differentiable_2025} further advanced this area with differentiable models that enable gradient-based optimization of effect parameters. This trend toward differentiable processing has facilitated integration with broader AI-assisted mixing and mastering systems, as shown by Koo et al. \cite{koo2023music}.
Despite these advances, fully ``black-box" neural audio effect modules often underperform compared to their traditional DSP counterparts, with even simple neural ``grey-box"  parameter controllers for conventional plugins showing comparable performance to end-to-end neural models~\cite{steinmetz2022style, steinmetz2023high}. 
% This highlights the continued value of traditional DSP expertise and suggests that hybrid approaches may offer the most promising path forward.

\subsection{Python Packages for Neural Audio Effect Processing}
The growing interest in neural audio effect modeling has spawned several Python packages aimed at facilitating research and development in this domain. These packages generally fall into two categories: differentiable implementations and interfaces to traditional audio plugins.
Among differentiable packages, DDSP (Differentiable DSP)~\cite{engel2020ddsp} provides differentiable implementations of common audio processing operations like filters and oscillators.
Other differentiable implementations include DiffMoog~\cite{uzrad2024diffmoog},  GraFX~\cite{lee2024grafx}, and PyNeuralFX~\cite{yeh2024pyneuralfx}, and NablaFX~\cite{comunita_nablafx_2025}.
In contrast to these differentiable implementations, a smaller number of packages provide interfaces to traditional audio plugins. Notably, \texttt{pedalboard} offers a simple Python interface to a limited set of built-in audio effects and VST plugins, and \cite{steinmetz2022style} developed custom interfaces for specific commercial plugins.
However, these existing interfaces typically support only a narrow range of plugin formats (especially regarding Windows-based plugins), and may lack control and support for complex routing topologies like sidechaining. 
% Additionally, most focus on single-effect processing rather than complete effect graphs, limiting their applicability to real-world audio production workflows.
% This gap between differentiable implementations and interfaces to traditional plugins highlights the need for more comprehensive solutions that can bridge research and industry practice. 
WildFX addresses this gap by providing a platform-agnostic interface to commercial plugins across multiple formats, supporting complex routing topologies, and enabling the generation of datasets that reflect actual industry practices rather than simplified approximations.

\section{Dataset \& Generation Pipeline}
% \textcolor{red}{Zachary[summary of this section]}

Here, we describe our methodology for building WildFX. As a data generation and processing pipeline, we first detail WildFX's dockerized deployment environment, followed by its interface protocol with the DAW \reaper{}. We then discuss the core data structure in WildFX, and how such objects are used for efficient data generation.

% Docker, \reaper{}, headless command-line only audio rendering system.

\subsection{Deployment Environment}
% We use Docker container to pack all the dependencies and related software to reach a end-to-end implementation. The dataset folder and linux plugins folder should be located outside of the container which mapped to the container, for the sake of saving storage. Three important software are already ready to go in the container built with our provided \texttt{Dockerfile}: (1) \reaper: a Digital Audio Workstation (DAW), (2) \texttt{Wine}: compatibility layer capable of running Windows applications on Linux, (3) \texttt{yabridge}: Windows audio plugins mounter.
% \texttt{Wine} and \texttt{yabridge} enable us to run most audio plugins in \texttt{.exe} format designed for Windows machines. REAPER can load all the plugins and render audio based on project metadata, which we will shed lights on it in Section~\ref{sec:pipeline}. The generated dataset can be accessed from outside of the container.

We utilize a Docker container to encapsulate all dependencies and software components required for end-to-end execution. To conserve storage, both the dataset directory and the Linux plugin folder are mounted from outside the container. Our provided \texttt{Dockerfile} sets up three key tools: (1) \reaper{}, a professional Digital Audio Workstation (DAW); (2) \texttt{Wine}, a compatibility layer for running Windows applications on Linux; and (3) \texttt{yabridge}, a compatibility layer for mounting Windows audio plugins. Together, \texttt{Wine} and \texttt{yabridge} allow REAPER to host \texttt{.exe}-format plugins designed for Windows on Linux servers. REAPER then renders audio based on project metadata, as detailed in Section~\ref{sec:pipeline}. All generated data are written to mounted directories and accessible outside the container.

% \end{minipage}%
% \hfill
\begin{wrapfigure}{r}{0.5\textwidth}
\vspace{-1em}
    \centering    \includegraphics[width=0.8\linewidth]{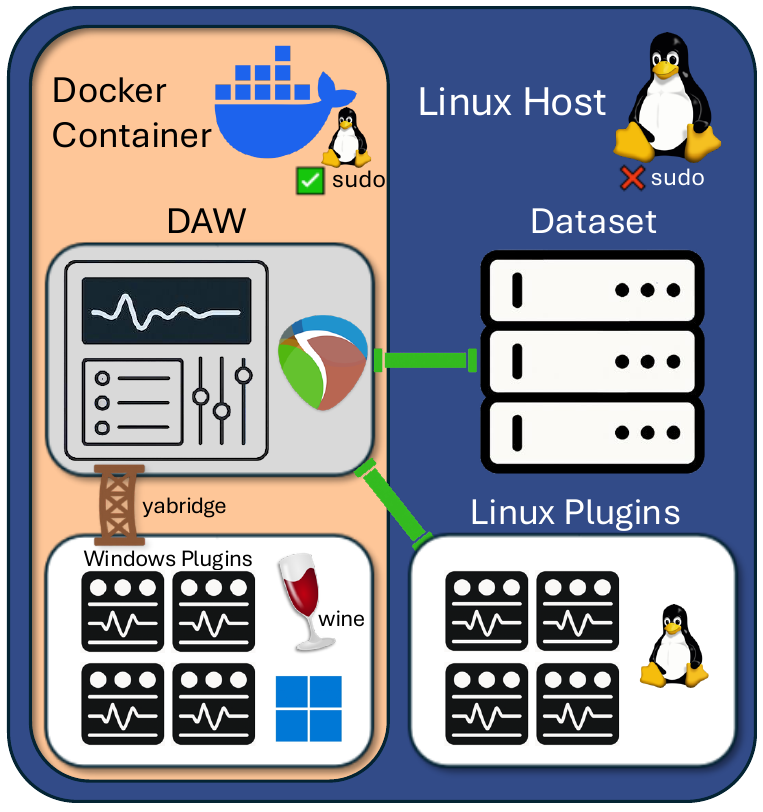}
    \captionof{figure}{Deployment Environment of \textbf{WildFX}}\label{fig:deploy}
    \vspace{-1em}
\end{wrapfigure}

% \texttt{Wine} and \texttt{yabridge} enable us to run most audio plugins in \texttt{.exe} format designed for Windows machines. REAPER can load all the plugins and render audio based on project metadata, which we will shed lights on it in Subsection~\ref{pipeline}. The generated dataset can be accessed from outside of the container.

% Docker is a platform for developing, shipping, and running applications in lightweight, isolated containers. With a single \texttt{Dockerfile} provided, one can build a container for our data generation pipeline with ease, without any trouble of running environment handling. This is very crucial to open-source projects. In our code repository, we prepared a config file \texttt{devcontainer.json} for the Dev Container Plugin in Visual Studio Code. We recommend to use it over manually attaching containers using command \texttt{docker run} as VS Code can automatically handle most of the file system mapping and authority issues. Besides, it provides easier checking as users can direcly look at the source code for better understanding not only limiting reading the help funciton, though all the functions are documented well.

% For a big part of users, they don't usually have the access to sudo command or root user authority. A large portion of audio plugins and \reaper{} require system-level dependencies, including \texttt{jackd} audio service which is essential to \reaper{}. [more argument about how important a linux audio processing system don't need sudo access]. The overall deployment architecture is pictured in Figure~\ref{fig:deploy}.

Docker is a platform for developing, shipping, and running applications in lightweight, isolated containers. Using the provided \texttt{Dockerfile}, users can easily build a self-contained environment for the WildFX data generation pipeline, eliminating the need for manual dependency management—a critical feature for reproducible open-source research. To further simplify setup, we include a \texttt{devcontainer.json} configuration compatible with Visual Studio Code's Dev Container extension. This approach is preferable to manual \texttt{docker run} workflows, as VS Code automates some file system mounting and permission handling, and offers a more transparent development experience by exposing the source code directly within the editor.

For many users, particularly in institutional or shared computing environments, \texttt{sudo} or root access is restricted. However, tools like \reaper{} and many commercial audio plugins require system-level dependencies, such as the \texttt{jackd} audio server, which is essential for proper DAW operation. WildFX addresses this challenge by encapsulating all necessary services within the container, allowing full Linux-based audio processing without elevated privileges. This design ensures accessibility across diverse user environments. The complete deployment architecture is illustrated in Figure~\ref{fig:deploy}.

% \subsection{DAW Control}
% We use python library \texttt{reapy} to control the Python API, \texttt{Reascipt}, provided by \reaper{}. \texttt{Reascript} itself allows users to control \reaper{} using Python, but it all depends on the Graphical UI of the system. \texttt{reapy} can be imported into any python environment and build connection with running \reaper{} instance. Utilizing the built-in data class \texttt{reapy.Project}, we can gain control of almost all basic features in \reaper{}, including selecting tracks, adding tracks, adding send (track signal routing). Part of the features related to the Project instance in \reaper{} that are not implemented by \texttt{reapy} are a number of important global controls, including project rendering, which can be realized by accessing the standard \texttt{Reascript} API in sub-module \texttt{reapy.reascript\_api}. Some important control like project rendering are not specifically implemented as API function. However, we can utilize the command sending API function \texttt{reapy.reascript\_api.Main\_OnCommand} to send specific command by ID. 

% Since we are controlling the \reaper{} client, which is originally designed to use with GUI, using API in headless mode, the client doesn't throw any error or provide a entrance for any inspection. The data generation pipeline would be highly unreliable without any manual checks of the rendered output. Therefore, we have validated every step we took is doing exactly what it's supposed to do with a liunx machine with GUI.
\subsection{DAW Control}
We use the Python library \texttt{reapy} to interface with \reaper{} through its scripting API, \texttt{ReaScript}. While \texttt{ReaScript} allows programmatic control of \reaper{} via Python, it typically relies on a graphical user interface. In contrast, \texttt{reapy} can be imported into any Python environment and establishes a connection to a running \reaper{} instance. Through its \texttt{Project} class, users can control essential DAW operations, including track creation, selection, and routing (e.g., adding sends).

However, some global project-level functions, such as rendering, are not fully exposed in \texttt{reapy}. These can be accessed via the lower-level API in \texttt{reapy.reascript\_api}, including through the general-purpose command interface \texttt{Main\_OnCommand}, which triggers REAPER actions by their command IDs.

Since \reaper{} is designed for interactive GUI use, operating it in headless mode via APIs provides no built-in error reporting or inspection tools. This limitation poses challenges for reliable automation. To ensure correctness, we validated every step of the pipeline manually on a Linux system with GUI support before deploying the system in headless environments.

% List of all commands are available online.\footnote{\href{https://raw.githubusercontent.com/Ultraschall/ultraschall-lua-api-for-reaper/Ultraschall-API-4.6/ultraschall_api/misc/misc_docs/Reaper-ActionList.txt}{https://raw.githubusercontent.com/Ultraschall/ultraschall-lua-api-for-reaper/Ultraschall-API-4.6/ultraschall\_api/misc/misc\_docs/Reaper-ActionList.txt}}
% \begin{figure}[ht]
    % \centering
    % \includegraphics[width=0.5\linewidth]{pics/deploy.pdf}
    % \caption{Deployment Environment of the Dataset Generation Pipeline}
    % \label{fig:deploy}
% \end{figure}

% \subsection{Intro to Processing Configurations & FX motifs}

% \subsubsection{Music Audios Componen}

% \subsubsection{Plugins}

\subsection{Data Structure}
% The WildFX pipeline introduces a carefully designed and robust data structure to facilitate detailed and comprehensive modeling of professional audio effect graphs. To maximize the accuracy and realism of generated datasets, WildFX employs a schema that is both hierarchical and highly configurable, ensuring compatibility with professional Digital Audio Workstations (DAWs) and a wide range of audio plugins. WildFX utilizes YAML and JSON file formats strategically to encode and manage distinct aspects of the audio effect graphs:
% \begin{itemize}
%     \item YAML Files: Primarily store the overarching structure of each audio mixing project. This includes the definition of audio input sources, their entry points into the effect graph, the explicit sequencing of audio effect chains (nodes), and the directed connections (edges) between these nodes.
%     \item JSON Files: Represent the detailed specifications and presets of individual audio effects. JSON files define the plugin names, parameter constraints, permissible parameter ranges, and specific preset configurations used in the audio effect nodes. This separation of structural definition (YAML) from plugin-specific details (JSON) simplifies maintenance, readability, and scalability.
% \end{itemize}
% We further provide a function to transfer the Project class we defined to \texttt{networkx} graph object which are demenstrated in \ref{sec:graph}.
The WildFX pipeline defines a robust and extensible data structure tailored for modeling professional audio effect graphs. To ensure compatibility with Digital Audio Workstations (DAWs) and commercial plugins, the schema is both hierarchical and modular. YAML files encode the high-level project structure, including audio input sources, their routing through effect chains, and the directed connections between nodes. JSON files specify plugin-level details such as parameter constraints, valid ranges, and preset configurations. This separation between structural and plugin-specific metadata improves clarity, reusability, and maintainability. Additionally, we provide utilities to convert the \texttt{Project} class into a \texttt{networkx} graph object, as discussed in Section~\ref{sec:graph}.

\subsubsection{Core Components}
The WildFX data schema comprises several interrelated Python classes, each representing a critical component of the audio effect graph:
\begin{itemize}
    \item \texttt{FXSetting}: Represents individual audio effects within an effect chain. Each \texttt{FXSetting} instance encapsulates the plugin's name (\texttt{fx\_name}), type (\texttt{fx\_type}), optional preset index (\texttt{preset\_index}), parameters (\texttt{params}), input/output channel counts (\texttt{n\_inputs}, \texttt{n\_outputs}), and optional sidechain configuration (\texttt{sidechain\_input}). Parameter validation is performed using constraints defined in the corresponding JSON preset files, ensuring DAW-compatible and realistic parameter settings. In Section~\ref{sec:layer}, we demonstrate how this structured representation facilitates efficient parallel processing.
    \item \texttt{ChainDefinition}: Encapsulates sequences of FXSetting objects into effect chains, defining how audio signals sequentially flow through multiple effects. Each chain specifies outgoing connections with gain information (\texttt{next\_chains}) to subsequent effect chains, enabling complex multi-path routing (e.g., parallel processing, sidechains). Empty chains (no processing for the input signal) are allowed for the completeness of all possible graph structure.
    \item \texttt{InputAudio}: Clearly defines audio input files, types and their entry points (input\_FxChain) into the graph, thus establishing explicit start points within the effect graph structure.
    \item \texttt{Project}: Represents the complete specification of an audio effect graph for a given mixing project. A Project instance aggregates multiple ChainDefinition instances, a set of InputAudio objects, and the final output audio path. Robust validation within the Project class ensures graph integrity, enforcing critical constraints such as acyclicity, valid indexing, correct sidechain routing, and logical consistency of audio inputs and outputs.
\end{itemize}

\subsubsection{Metadata Sample}
\begin{minipage}[t]{0.48\textwidth}
\centering
\textbf{YAML Project Metadata Example}
{\small \begin{verbatim}
FxChains:
  - FxChain:
      - fx_name: "VST3: 3 Band EQ"
        fx_type: "eq"
        preset_index: 2
        params: []
        sidechain_input: null
    next_chains:
      1: 1
  - FxChain: []

input_audios:
  - audio_path: "vocals.wav"
    audio_type: "vocal"
    input_FxChain: 0

output_audio: "mixed_output.wav"
customized: true
\end{verbatim}}
\end{minipage}
\hfill
\begin{minipage}[t]{0.48\textwidth}
\centering
\textbf{JSON Plugin Preset Example}
{\small \begin{verbatim}
{
  "fx_name": "VST3: 3 Band EQ",
  "fx_type": "eq",
  "n_inputs": 2,
  "n_outputs": 2,
  "valid_params": {
    "Low": [0.0, 0.01, ..., 1.0],
    "Mid": [0.0, 0.01, ..., 1.0],
    "High": [0.0, 0.01, ..., 1.0]
  },
  "presets": [
    [null, null, null, 0.12, 0.69, 0.21],
    [null, null, null, 0.72, 0.63, 0.09],
    [null, null, null, 0.05, 0.00, 0.28]
  ]
}
\end{verbatim}}
\end{minipage}
\begin{wrapfigure}{r}{0.5\textwidth}
\vspace{-1em}
    \centering    \includegraphics[width=0.8\linewidth]{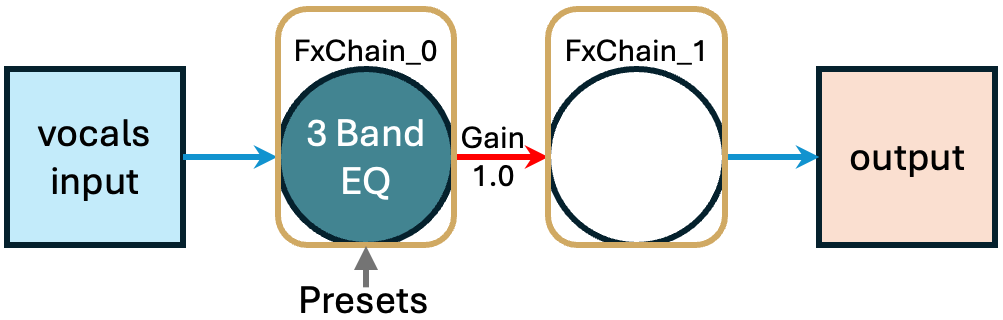}
    \captionof{figure}{Mixing Graph with the Provided Sample}\label{fig:example}
    \vspace{-1em}
\end{wrapfigure}

% \begin{figure}[ht]
%     \centering
%     \includegraphics[width=0.5\linewidth]{pics/example.png}
%     \caption{Minimal example of mixing graph corresponding to the metadata samples. The circle on the right side represents a empty \texttt{FxChain}.}
%     \label{fig:example}
% \end{figure}
Note that in the provided project metadata example, the \texttt{params} field is empty, as the plugin uses the preset indexed by 2 from its corresponding JSON preset file. All plugin parameters are sampled from the discretized set defined in \texttt{valid\_params}. A \texttt{null} value in the JSON file indicates use of the plugin’s default setting. Additionally, the values in \texttt{next\_chains} represent gain in linear amplitude and are not rescaled to perceptual units (e.g., dB). Figure~\ref{fig:example} illustrates the corresponding graph structure.

% Note that in the provided project metadata sample, \texttt{params} is a empty list, since it can read the preset of index 2 in its corresponding JSON plugin preset file. All the parameters of plugins would be sampled from the discretized valid parameter set \texttt{valid\_params} in JSON plugin preset. In the JSON preset file, null value represent accepting plugins' default value. Note that the values of \texttt{next\_chains} is gain (waveform amplitude) in linear scale haven't been rescale to adapty human perception using log scale (dB). Figure~\ref{fig:example} shows the corresponding graph of the minimal example.

\subsubsection{Graph Features Designation}\label{sec:graph}
The heterogeneous graph data are stored as Python pickled \texttt{networkx} graph objects for efficient loading. For audio input nodes, it has three features: \texttt{type='audio'}, \texttt{label} and \texttt{instance}, where \texttt{instance} marked the exact source of the audio and \texttt{label} could be defined for specific tasks. AFx plugin nodes, whose \texttt{type} should be \texttt{'fx'}, have an additional feature \texttt{params} stored as a Python dictionary holding all parameters specific to each AFx plugin instance.

The \texttt{type} and \texttt{label} features also exist in edges along with a \texttt{gain} feature. All available values are: \texttt{type=\{'send\_signal', 'split\_signal'\}}, \texttt{label=\{'main', 'control'}\}. For example, the feature of a splitter's outgoing edge would be \texttt{\{type='split\_signal', label='main'\}}, and the feature of the incoming edge for sidechain/control signal would be \texttt{\{type='send\_signal', label='control'\}}. \texttt{gain} are all in linear scale.

\subsubsection{Data Structure Restrictions}\label{sec:datarestriction}
For the ease of batch rendering in a headless client, we impose the following restrictions to the data structure metadata files.
\begin{enumerate}
    \item \textbf{Graph Acyclicity:} The audio effect graph must be a Directed Acyclic Graph (DAG).
    % \item \textbf{Valid Chain Indices:} All references to chain indices (e.g., in \texttt{next\_chains}, \texttt{input\_FxChain}, and \texttt{sidechain\_input}) must correspond to valid chain indices.
    % \item \textbf{Defined Inputs:} Chains designated as input entry points via \texttt{input\_audios} must have no predecessors.
    \item \textbf{I/O:} There must be exactly one final output chain with no outgoing connections and at less one input soruce.
    % \item \textbf{FX Type Compliance:} Every effect type (\texttt{fx\_type}) must be within the predefined allowed set.
    \item \textbf{Splitter Rules:} (1) Splitters must be positioned as the last effect in a chain. (2) Chains with multiple outgoing connections (\texttt{next\_chains}) must have a splitter as the last effect. (3) The final output chain cannot end with a splitter.
        % \end{itemize}
    \item \textbf{Sidechain Constraints:} (1) Only one sidechain-enabled plugin per chain is allowed. (2) Sidechain source chains must not originate from splitter outputs. (3) Sidechain routing can only happen within the chains inside the same project and same layer.
        % \end{itemize}
    % \item \textbf{Parameter Validity:} Effect parameters must match allowed values and ranges defined in their corresponding JSON preset files.
    % \item \textbf{Input and Output Channels:} \texttt{n\_inputs} and \texttt{n\_outputs} must match values specified in the corresponding preset JSON.
    % \item \textbf{Non-Negative Gains:} Connection weights (gain values) defined in \texttt{next\_chains} must be non-negative floats.
\end{enumerate}
Note that the third restriction of sidechain does not forbid the recurrent sidechain routing, i.e., tracks in the same layer in the same project is requiring each other as control signal. Although it is not technically possible when using DAW in traditional method, it could happen in purely random graph generation. Our restriction of Graph Acyclicity actaully rule this situation out. By applying these restrictions, we guarantee that all rendering tasks defined by each \texttt{FxChain} data class can be done in one execution. Later in Section~\ref{sec:theory}, we will show that those restrictions, especially on Splitters and Sidechain, will not limit the abundance of graph we can generate in the pipeline.

\subsection{Generation Pipeline}\label{sec:pipeline}
% The typical workflow of WildFX is depicted in Figure~\ref{fig:workflow}. After installing the audio effect plugins we want to use and starting \reaper{}, we run a \texttt{.lua} script to get a comprehensive list of all plugins that \reaper{} can detect. From the list, we can choose the plugins that we want to use to generate presets by \texttt{gen\_presets.py}. Provided all the JSON preset files, we can run \texttt{gen\_projects.py} to generate project specific YAML metadata. Then the main function would start to render audio, automatically handling all the arguments in the project metadata.

\begin{figure}[ht]
    \centering
    \includegraphics[width=1\linewidth]{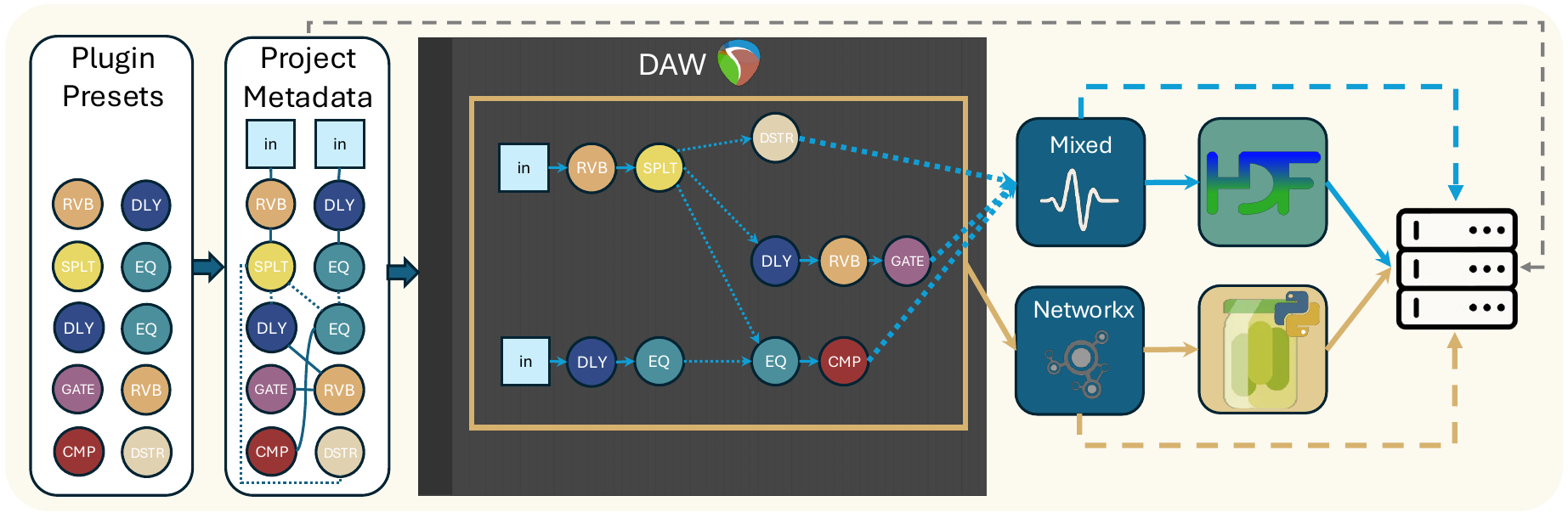}
    \caption{Overview of the \textbf{WildFX} workflow. Users begin by specifying plugin names to generate corresponding presets. These presets are then used to synthesize project metadata, which defines audio effect graphs. The metadata is rendered using \reaper{} in a headless environment, after which the output can be exported in the desired format (e.g., waveform, compressed HDF5, YAML graph specification or pickled \texttt{networkx} objects). External dashed lines indicate optional data storage.}
    % \caption{Workflow of \textbf{WildFX}: generate plugin presets by specifying the plugin names, generate project metadata with the presets generated, process with \reaper{}, and finally choose the desired data format.}
    \label{fig:workflow}
\end{figure}

The typical workflow of the WildFX pipeline is illustrated in Figure~\ref{fig:workflow}. After installing the desired audio effect plugins and launching \reaper{}, a bundled \texttt{.lua} script is executed to retrieve a complete inventory of all plugins recognized by the DAW. From this inventory, a subset of plugins can be selected for preset generation using \texttt{gen\_presets.py}, which creates structured JSON files defining valid parameter spaces and example presets. Once these JSON preset files are available, \texttt{gen\_projects.py} can be used to generate project-specific YAML metadata, encoding both the audio routing topology and plugin configurations. The topological structure is generated layer-by-layer with the algorithm we use in layer-based batch processing \ref{fig:layer}. Finally, the main rendering engine processes this metadata to render the audio in a fully automated manner, respecting plugin order, signal flow, and sidechain routing as specified in the project graphs.

\subsubsection{Pipeline Interface}
To support large-scale preset generation from real-world audio plugins, our system provides a modular command-line interface with key functionalities outlined below:

In the preset generation part:
\begin{itemize}
    \item \textbf{Plugin Selection.} Users can specify plugins via (1) \texttt{--plugin-name \{NAME TYPE\}} to target specific plugins, (2) \texttt{--plugin-list} to load a batch from a CSV, and (3) \texttt{--use-reduced-set} or \texttt{--use-full-set} to quickly sample canonical plugin sets.

    \item \textbf{Parameter Sampling.} The system would dicretize the range of all the interested parameters with suitable distribution to specific parameter's nature to avoid the sampling heterogeneity across a large set of distinct parameters.

    \item \textbf{Cluster-Based Validation.} Optionally enabled via \texttt{--validate\_generation}, the system renders audio from sampled parameter sets and clusters them in MFCC space using KMeans to select representative presets.
\end{itemize}

To generate structured multi-track mixing projects with realistic audio effect graphs, our system provides a configurable interface that supports the following key functionalities:

\begin{itemize}
    \item \textbf{Dataset-Aware Stem Extraction.} The user can define customized logic to extract stems and instrument labels (e.g., \texttt{Bass}, \texttt{Guitar}) via \texttt{--dataset-name} and project folder parsing. For example, our interface supports original dataset structure to be similar to the Slakh2100 Dataset\cite{manilow2019cuttingmusicsourceseparation}.

    \item \textbf{Topology Synthesis.} The pipeline builds directed acyclic graph (DAG) topologies with configurable complexity (\texttt{--complexity}), number of FX chains (\texttt{--min-chains}, \texttt{--max-chains}), number of input stems (\texttt{--min-stems}, \texttt{--max-stems}), and controlled probability of sidechains and splitters.

    % \item \textbf{Plugin Assignment.} Available FX plugins are loaded from a preset directory and sampled per chain with optional sidechain insertion. Chains needing splitters automatically insert plugins with matching band counts.

    \item \textbf{Parametric Control.} Each chain's depth is sampled from a user-defined categorical distribution (e.g., \texttt{--chain-depth 0.1,0.6,0.3}) to model varying processing complexity across chains.

    \item \textbf{Variability Support.} The \texttt{--variable-density} flag enables randomization of graph density, depth distribution, and probabilities per project, emulating diverse real-world mixing styles.

    % \item \textbf{Batch Project Generation.} The script supports generating large numbers of projects (\texttt{--num-projects}), with automatic resampling over existing folders to meet the target count. Output is serialized in validated YAML format using the shared data schema.
\end{itemize}

In main function, users can choose the data saving mode to be in human-readable format: \texttt{.wav} and \texttt{.yaml}, or training-ready format: \texttt{HDF5} for audio and \texttt{pickle} files of networkx graph objects, for efficient I/O management in training deep learning models.

\subsubsection{Layer-based Multi-Project Batch Processing}\label{sec:layer}
\begin{figure}[ht]
    \centering
    \includegraphics[width=0.8\linewidth]{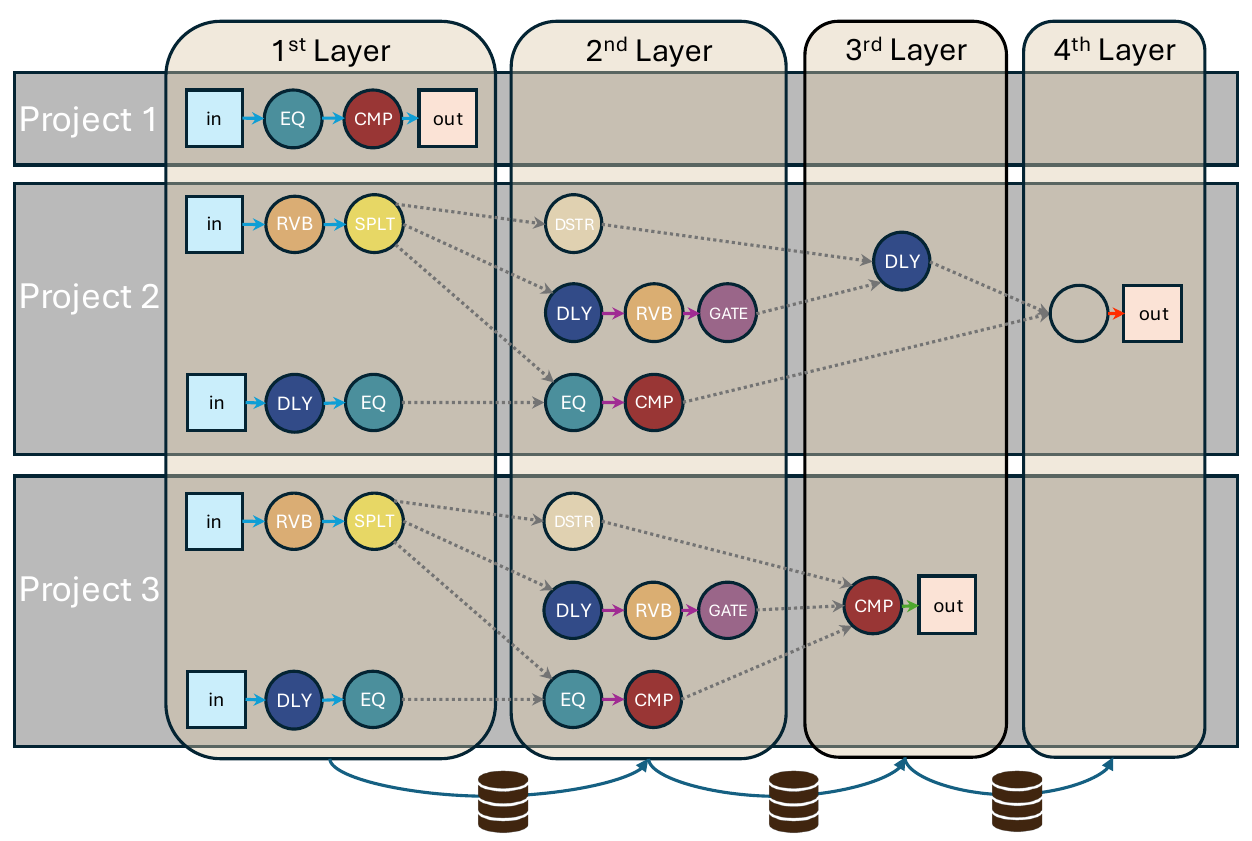}
    \caption{Layer-based Multi-Project Batch Processing: the 4 layers across 3 projects are determined by the Kahn's topological sorting algorithm treating AFx processing paths as hyper-nodes. The directed edges of mixing path which are treated as hyper-nodes are marked as the same color in each layer. The edges across layers are marked as grey dashed arrows. Each time the DAW with batch process all the mixing path within the same current layer, store all the output requried by the next layer, and then repeat this process recursively.}
    \label{fig:layer}
\end{figure}

The minimal processing unit in DAWs is a single path of AFx. Therefore, we need divide the whole processing graph into chunks only containing paths. To ensure efficient and dependency-aware rendering of complex audio effect graphs, WildFX employs a layer-based execution strategy inspired by Kahn's algorithm for topological sorting. In this scheme, each project is treated as a directed acyclic graph (DAG), where:

\begin{itemize}
    \item Each node corresponds to an \texttt{FXChain} (in the graph data we provide, each node represents a plugin), representing a sequence of AFx plugins applied.
    \item Directed edges encode audio signal flow, including main connections and sidechain routings, between chains.
\end{itemize}
This is the major reason why we adopt the \texttt{FxChain} data structure. Mixing multiple tracks directly within the DAW is inefficient, particularly in a headless rendering environment. To address this, WildFX performs all mixing operations in advance by summing the raw audio waveforms within Python prior to effect processing at each layer. This strategy enables straightforward parallelization and significantly reduces the overhead associated with in-DAW signal routing.
% Mixing multiple tracks are not efficient using DAW, especially with our headless environment. Therefore, we do all the mixing before the processing at each layer by adding up the raw audio data read into Python, which is parallelizable.

At runtime, the system processes nodes in dependency-respecting layers: (1) Nodes with zero unresolved dependencies across all projects (i.e., in-degree zero) form the current layer, (2) These nodes are batch-processed in parallel, including audio mixing, plugin parameterization, and rendering via \reaper. (3) Upon successful execution, their outputs are stored, and in-degrees of successor nodes are decremented. (4) Nodes whose in-degree reaches zero become eligible for the next layer.

This approach depicted in Figure~\ref{fig:layer} ensures deterministic, deadlock-free scheduling while supporting advanced DSP behaviors, including: \textbf{(1) Splitter-aware routing}: handling multi-output plugins that branch signals into parallel chains, \textbf{(2) Sidechain synchronization}: ensuring sidechain sources and consumers reside in the same layer, \textbf{(3) Batch-aware optimization}: grouping tasks to maximize CPU/GPU utilization under resource constraints.

By embedding topological constraints directly into the processing loop, WildFX achieves scalable and correct simulation of arbitrarily structured audio effect graphs.

\subsection{Mixing Graph Completeness}\label{sec:theory}
We can define all the audio mixing graphs to be a DAG with only one node having the outdegree of 1, and at least one node having indegree of 1. If multiple output is needed, we can separate them by the number of outputs by simply duplicating the repeated or shared AFx settings along the processing procedure in different output paths. 

In Section~\ref{sec:datarestriction}, we also restricted that either only one AFx plugin with sidechain enabled or only one splitter at the end of a \texttt{FxChain} is allowed. It is very natural to think that this might cause some specific mixing structure to be infeasible for our pipeline. However, we argue that with the permission to empty chains, our data structure can actually represent all possible mixing graph structures.
For the case when splitters and plugins enabling sidechain signals are present in the same \texttt{FxChain} and multiple plugins requiring control signals from sidechain, we can separate them into several \texttt{FxChain} only containing one such plugin. For chains that having multiples or intermediate splitters, we can also separate the chain into several subchains according to the location of the splitters, ensuring splitter is always in the end to fit in our universal concurrent single track processing. 

We ruled out the situation when a plugin requires control signal from a split stem from a splitter. Because in this case, the split stem created by receiving signal from sends from other track would send it channel to new tracks. If allowing this case, out processing batch could form complex nested sending configuration which is hard to implement. For this case, we can first split the chain and move the latter part containing the plugin asking for control signal to next layer, and add a pseudo layer containing no processing (empty chain) after the split stem asked to be control signal. This two-layer separation is equivalent to its one-layer origin.

% two chains both have a plugin enabling sidechain control, and requiring each other's output as control signal. In the sense of main signal transmitting, they should be in the same layer. But since they require each other's output, they cannot be in the same layer. For this case, we can first separate one of those chains, named as A, moving its part containing the sidechain plugin to the next layer. At the same time we add a pseudo layer containing no processing (empty chain) right after chain B. In the subsequent layer processing, the remaining part of A can actually use the processed output from chain B as control signal, which is equivalent to the original signal routing. 

\section{Experiments}
\subsection{Dataset}
% Utilizing \textbf{WildFX} we generated two dataset named as shallow dataset and deep based on the Slakh2100 dataset with different settings. Each training set and validation set contains 5,000 and 270 multi-track projects, respectively. The plugins we used to generate the dataset is listed in Table~\ref{tab:plugin-inventory}. For both of the dataset, we utilized \texttt{--variable-density}, sidechain probability of 0.2, splitter probability of 0.1. The deep dataset can sample stems from \texttt{\{piano, guitar, bass, drums\}}, while the shallow dataset is limited to \texttt{\{guitar, bass\}}. Distinct parameters used to generate the dataset are given in Table~\ref{tab:config-params}.

Using the \textbf{WildFX} pipeline, we generated two datasets—referred to as the \textit{shallow} and \textit{deep} datasets—derived from the Slakh2100 corpus, each configured with distinct structural and signal processing parameters. Both datasets consist of 5,000 training projects and 270 validation projects, each representing a multi-track audio mixing session. The audio effect plugins employed in the dataset generation are listed in Table~\ref{tab:plugin-inventory}. For both configurations, we enabled \texttt{--variable-density}, with a sidechain probability of 0.2 and a splitter probability of 0.1. The deep dataset samples instrument stems from a broader pool, including \texttt{\{piano, guitar, bass, drums\}}, whereas the shallow dataset is restricted to \texttt{\{guitar, bass\}}. The specific generation parameters for each dataset are detailed in Table~\ref{tab:config-params}.

% The average processing times for the shallow and deep datasets are approximately 12 seconds and 15 seconds per project, when executed on a machine with dummy \texttt{jackd} server with 64 CPU cores; 3.5 seconds and 5 seconds on a machine with alsa \texttt{jackd} server with 64 CPU cores.
With a 64-core CPU, average processing times for the shallow and deep datasets were 12 and 15 seconds under a dummy \texttt{jackd} server, compared to 3.5 and 5 seconds with an alsa \texttt{jackd} server.

\begin{table*}[ht]
\raggedright
\begin{minipage}[t]{0.45\textwidth}
\raggedright
\caption{Plugin Inventory}
\begin{tabular}{lcc}
\toprule
\textbf{AFx Plugin} & \textbf{Format} & \textbf{Type} \\
\midrule
3 Band EQ & VST3 & EQ \\
3-Band Splitter & JS & Splitter \\
Samurai Delay & VST3 & Delay \\
Schroeder & VST3 & Reverb \\
ZamCompX2 & VST3 & Compressor \\
\bottomrule
\end{tabular}
\label{tab:plugin-inventory}
\end{minipage}
\hfill
\begin{minipage}[t]{0.5\textwidth}
\raggedleft
\caption{Dataset Configuration Parameters}
\begin{tabular}{lcc}
\toprule
\textbf{Parameter} & \textbf{Shallow} & \textbf{Deep} \\
\midrule
\makecell[l]{\texttt{--min/max} \\ \texttt{chains}} & 3, 5 & 3, 10 \\
\makecell[l]{\texttt{--min/max} \\ \texttt{stems}} & 1, 2 & 1, 4 \\
\makecell[l]{\texttt{--chain-depth} \\ \texttt{distribution}} & [.1, .7, .2] & [.1, .3, .4, .2] \\
\bottomrule
\end{tabular}
\label{tab:config-params}
\end{minipage}
\end{table*}

\subsection{Blind Estimation of Mixing Graphs}
% We implemented the audio mixing graph blind estimation methods in \cite{lee_blind_2023}. The method first first encodes the reference y into latent vectors z, which should contain the necessary information to estimate the graph. Then, from the latent z, we reconstruct the graph in two stages, which resemble the synthetic data generation procedure; we first decode the prototype Gˆ0 autoregressively, then estimate the remaining parameters pˆ. Utilizing two available models, autoencoding and decoding, we get results from 4 different setting. 

We implemented the audio mixing graph blind estimation method proposed by Lee et al.~\cite{lee_blind_2023}. The approach begins by encoding the reference audio signal $y$ into a latent representation $z$, which is expected to capture the information necessary to infer the underlying processing graph. The graph is then reconstructed in two stages, mirroring the synthetic data generation pipeline: first, a prototype graph $\hat{G}_0$ is decoded autoregressively to recover the structural topology; subsequently, the remaining parameters $\hat{p}$ are estimated. By leveraging two available model configurations. autoencoding and prototype decoding, we evaluate performance under four distinct experimental settings.

\begin{table*}[ht]
\centering
\caption{Comparison of performance metrics across dataset configuration and processing settings. \textbf{PT Loss}: Prototype Decoding Loss. \textbf{PR Loss}: Parameter Loss}
\resizebox{\textwidth}{!}{
\begin{tabular}{lccccc}
\toprule
\textbf{Setting} & \textbf{PT Loss} & \textbf{PR Loss} & \textbf{Gain Loss} & \textbf{Edge Error Rate} & \textbf{Node Error Rate} \\
\midrule
Shallow + Autoencoding & \textbf{2.335} & 2.121 & \textbf{1.101} & 0.087 & \textbf{0.568} \\
Shallow + Decoding     & 2.511 & 2.096 & 1.120 & 0.093 & 0.625 \\
Deep + Autoencoding   & 2.872 & \textbf{2.089} & 1.417 & \textbf{0.070} & 0.625 \\
Deep + Decoding       & 2.927 & 2.448 & 1.510 & 0.094 & 0.690 \\
\bottomrule
\end{tabular}
}\label{tab:exp}
\end{table*}
% We trained the model with a AdamW [27] optimizer, a linear learning rate scheduler with 3e-4 peak learning rate, 50k warmup steps, 200k total training steps, and batch size of 32.

% The prototype loss, parameter loss and gain loss measure the distance between ground-truth and predicted mixing graph structure, plugin parameters and gain. The method give pretty good results on edge prediction, but struggle to predict nodes. This lack of structural awareness is also reflected in the prototype loss. Gain loss, which is related to edge attributes, is signaficantly better. Overall, autoencoding methods are performing signaficantly better than proposed prototype decoding method in all 5 metrics.

% \subsection{Generation Rate}
We trained the model using the AdamW~\cite{loshchilov2019decoupledweightdecayregularization} optimizer with a peak learning rate of 3e-4, a linear learning rate scheduler, 1k warmup steps, 3k total training steps, and a batch size of 32. Table~\ref{tab:exp} reports results across various dataset configurations and decoding strategies. We evaluate five metrics:
\begin{itemize}
    \item \textbf{Prototype (PT) Loss}: Measures the structural distance between predicted and ground-truth graph topologies.
    \item \textbf{Parameter (PR) Loss}: Captures errors in plugin parameter prediction.
    \item \textbf{Gain Loss}: Quantifies the deviation of predicted edge weights (gain values).
    \item \textbf{Edge Error Rate}: Measures the proportion of incorrectly predicted connections between nodes.
    \item \textbf{Node Error Rate}: Indicates incorrect prediction of node types and configurations.
\end{itemize}

Autoencoding models consistently outperform prototype-decoding counterparts across all metrics. In particular, edge-related metrics (e.g., Gain Loss and Edge Error Rate) show lower error compared to Node Error Rate, indicating the model’s stronger capability in recovering edge attributes than in identifying node types. The relatively high prototype loss further reflects this limitation in structural reasoning.

\section{Limitations \& Discussions}

Our results fall short of those reported in the original work~\cite{lee_blind_2023}, which we attribute to two key differences. First, their implementation used simplified plugins with minimal functionality, reducing task complexity. In contrast, we employ real-world plugins with more diverse behaviors. Second, our training set is approximately 100 times smaller due to our current resource constraints. Despite this, the reproduced model achieves competitive results and remains the state of the art for this task.

These findings underscore the challenge of realistic mixing graph estimation and highlight significant room for improvement. Our dataset establishes a strong benchmark for future research in this domain.

\bibliographystyle{plain}
\bibliography{main}

\end{document}